\documentstyle[psfig,twocolumn,aps]{revtex}
\newcommand{\tab}{TaB$_2$}
\newcommand{\mgb}{MgB$_2$}
\begin{document}
\draft
\wideabs{ \title{ Electronic structure and weak electron-phonon
coupling in \tab} 
\author{H.\ Rosner$^*$ and W.E.\ Pickett}
\address{Department of Physics, University of California, Davis, CA
95616, USA}
\author{S.-L.\ Drechsler, A.\ Handstein, G.\ Behr, G.\ Fuchs, K.\ Nenkov,  
K.-H.\ M\"uller, and H.\ Eschrig}
\address{ Institut f\"ur Festk\"orper- und Werkstofforschung Dresden e.V.,
Postfach 270116, D-01171 Dresden, Germany}
\date{\today}
\maketitle
\begin{abstract}
We present electronic structure calculations together with
resistivity, susceptibility, and specific heat measurements for \tab\
to search for the recently contradictorily reported superconductivity
and to study related normal state properties.  We ascribe the absence
of superconductivity down to 1.5 K for our \tab\ samples to the
generally weak electron phonon coupling derived from comparison of the
calculated and measured specific heat constants. For the E$_{2g}$ and
the B$_{1g}$ $\Gamma$ point phonons we derive from the calculated
deformation potentials very small electron phonon couplings for these
modes, opposite to the strong coupling of the E$_{2g}$ mode in
MgB$_2$, probably responsible for its high $T_c$.  In comparison to
MgB$_2$, we discuss the origin of the quite different features in the
density of states and of the Fermi surfaces. The differences are
mainly due to the strong hybridization between Ta 5$d$ and B 2$p$
states outside the hexagonal basis plane.
\end{abstract}
}
\narrowtext

\section{introduction}

The recent discovery of superconductivity in MgB$_2$\cite{akimitsu00}
has initiated an immediate broad research activity due to the
surprisingly high transition temperature $T_c\sim $ 40 K in a
seemingly ordinary $s-p$ metal.  Investigation of related $s-p$
diborides MB$_2$ (M = Li, Be, Al) \cite{slusky,felner,zhao} and a
series of isostructural transition metal diborides (M = Sc, Ti, Zr,
Hf, V, Ta, Cr, Mo, Nb) \cite{samsonov,nbb2_1,nbb2_2} has shown that
only few of them seem to be superconducting, and then only at very low
temperatures. Since the understanding of the pairing mechanism in
MgB$_2$ is still in its early stages, a study of the isomorphic
compounds with low transition temperatures or with absence of
superconductivity, respectively, could be helpful in clarifying the
expected very substantial electronic differences.

Contradictory reports about superconductivity in \tab\ have appeared;
one found superconductivity at $T_c \sim$ 9.5
K,\cite{kaczorowski1,kaczorowski2} while another found no
superconductivity down to 4.4 K.\cite{gasparov} TaB$_2$ is isovalent
with NbB$_2$, where the occurrence of superconductivity, or at least
the value of T$_c$, is likewise uncertain.  NbB$_2$ was reported to be
superconducting at 3.87 K,\cite{nbb2_1} superconducting only at 0.62
K,\cite{nbb2_2} and not superconducting above 0.37 K.\cite{gasparov}
Clearly there is sample dependence for both \tab\ and NbB$_2$, and no
doubt for many other diborides as well.

In an attempt to begin to settle some of these discrepancies, we
address TaB$_2$ specifically in this paper with a joint theoretical
(Secs.\ II A and III A) and experimental (Secs.\ II B, III B)
investigation.  In Sec.\ III we provide an analysis of its electronic
structure, and contrast it with that of \mgb.  We also present
specific heat, susceptibility as well as resistivity data for two
almost single phase samples.  One of them exhibited a significant
amount of boron vacancies.

\section{Methods}

\subsection{Band structure calculations}

We calculated the electronic structure of \tab\ in the hexagonal
space group (SG) P6$_3$/$mmc$ (No.~191) with the lattice constants $a$
= 3.082 \AA\ and $c$ = 3.243 \AA .\cite{kaczorowski1} The frozen
phonon calculations for the E$_{2g}$ and B$_{1g}$ modes where done in
the orthorhombic SG C$mmm$ (No.~65) and in the trigonal SG P${\bar
3}m1$ (No.~164), respectively.

Our band structure calculations were performed using the
full-potential nonorthogonal local-orbital minimum-basis scheme (FPLO)
\cite{koepernik99} within the local density approximation (LDA). In
these scalar relativistic calculations we used the exchange and
correlation potential of Perdew and Zunger.\cite{perdew81}
Ta $5s$, $5p$,$6s$, $6p$, 5$d$ states and B 2$s$, 2$p$, 3$d$, were
chosen as minimum basis set for the valence states. All lower lying
states were treated as core states. The inclusion of the relatively
extended Ta $5s$, $5p$ semi-core states as band states was done
because of the considerable overlap of these states on nearest
neighbors.  This overlap would be neglected if they were treated as
core states in our FPLO scheme. Accounting for this overlap is of
importance especially for the calculations of phonon frequencies that
we report.
B 3$d$ states were added to allow for boron polarizability. The
spatial extension of the basis orbitals, controlled by a confining
potential \cite{eschrig89} $(r/r_0)^4$, was optimized to minimize the
total energy.
The self-consistent potentials were carried out on a $k$-mesh of 20
$k$-points in each direction of the Brillouin zone, which corresponds
to 481, 784 and 1221 in irreducible part for P6$_3/mmc$, P${\bar
3}m1$, and C$mmm$ SG, respectively.
Our FPLO approach has been proved by recent comparison to FLAPW
results in many cases to have an absolute accuracy of 1mHartree/atom
for the total energy and a much higher accuracy for the relative total
energy changes.

\subsection{Sample preparation and measurements}

Two different TaB$_2$ samples, one with near stoichiometric and one
with boron deficient compositions, were prepared starting from the
pure elements Ta ($>$ 99.9\%) and B ($>$ 99.9\%) by arc melting
furnace under purified argon atmosphere.
To ensure a better homogeneity the samples were turned several
times. The phase content was checked by X-ray diffraction (XRD) using
Co-K$\alpha$ radiation.

The samples contain the hexagonal AlB$_2$ phase mainly, 95\% in
sample No.~1 and 97\% in sample No.~2. In both samples small amounts
($<$ 5\%) of a second phase were found which are B for sample No.~1
and Ta$_3$B$_4$ for sample No.~2. The compositions of the main phases
measured by electron probe microanalysis in the WDX modus showed a
composition of about TaB$_{2.03}$ (sample No.~1) and compositions of
TaB$_{1.29}$ (sample No.~2).

Furthermore, the XRD resulted in quite different lattice parameters
for the two  samples ranging from $a$ = 3.098$\pm$0.002 nm and
$c$ = 3.224$\pm$0.003 \AA\ of the B-deficient sample to $a$ = 3.067 $\pm$
0.002 \AA\ and $c$ = 3.286 $\pm$0.006 \AA\ of the B-rich sample.

The specific heat of both TaB$_2$ samples was measured in the
temperature range between 2 and 16 K using a Quantum Design PPMS
relaxation calorimeter.  The addenda which were determined in a
separate run were subtracted in order to obtain the specific heat data
for TaB$_2$.

Magnetization measurements have been performed using a Quantum
Design-SQUID magnetometer in the temperature range down to 1.8 K.
Resistivity measurements down to 1.5 K have been performed using the
standard four point method.

\section{Results and discussion}

\subsection{Theoretical results}
In Figure 1 we display the total as well as the atom decomposed
density of states (DOS) of TaB$_2$.  The B $2p$ and Ta $5d$ states
share almost equally in the occupied valence bands in the region -10
to -2 eV (the Fermi level $E=\varepsilon_F$ is taken as the zero of
energy).
Our DOS is in agreement with that of P.P.\ Singh\cite{ppsingh}, which
we became aware right after the completion of our study.

A striking difference in comparison to \mgb\ is the dominating
contribution of Ta $5d$ states to the DOS at Fermi level, which
contribute about 70\% of the total DOS; in \mgb\ the DOS at Fermi
level is dominated by B-$2p$ states (see Figure 2 for comparison
\cite{remark1}). Although a rigid band picture is very limited
in this case, a valence analysis shows that due to the
3 additional valence electrons of Ta with respect to Mg, the Fermi
level has shifted from the bonding B states below the hybridization
gap in \mgb\ to the anti-bonding states above this gap in \tab .
From the very similar shape of the partial B and Ta DOS, a strong
hybridization between B $2p$ and Ta $5d$ states is obvious, mentioned
below in more detail.

\begin{figure}[bt]
\psfig{figure=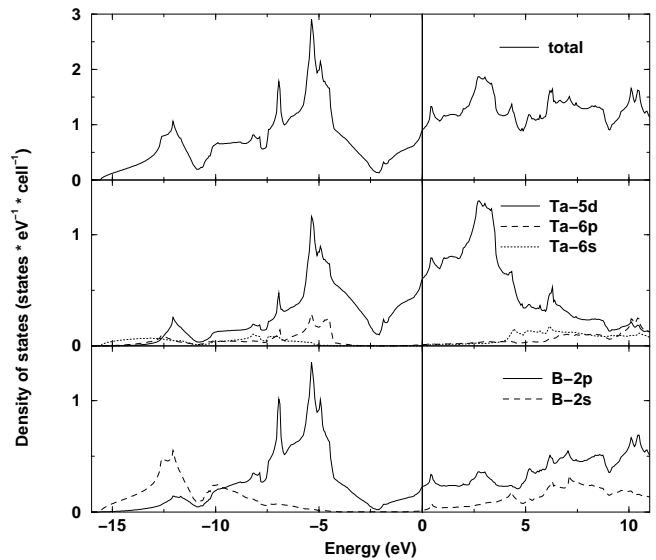,angle=-90,width=8.5cm}
\vspace{3mm}
\caption{ Total DOS and partial DOS of \tab . The upper panel gives
the total DOS, the middle and the lower panels show the contribution of Ta
and B, respectively. The Fermi level is at zero energy.}
\label{dos}
\end{figure}
The calculated value of
the density of states at the Fermi level $N(\varepsilon_F)$ is
slightly higher for \tab\ ($N(\varepsilon_F)$ = 0.91
states/(eV$\times$cell)) than for \mgb\ ($N(\varepsilon_F)$ = 0.71
states/eV$\times$cell), in agreement with
Refs.~\onlinecite{pickett,kortus}). This corresponds to a bare
specific heat coefficient $\gamma_0$ = 2.14 mJ/(mole$\times$K$^2$) for
\tab .
\begin{figure}[bt]
\psfig{figure=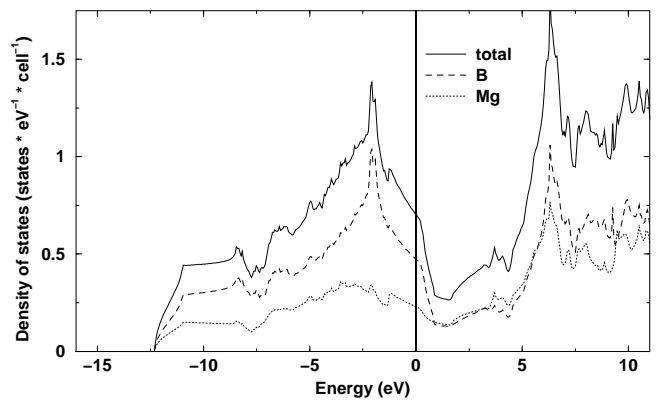,angle=-90,width=8.5cm}
\vspace{3mm}
\caption{ Total DOS and partial DOS of \mgb . The partial DOS are
calculated by projection to the minimum orbital basis. The Fermi level
is at zero energy.}
\label{dosmg}
\end{figure}
Figure 3 shows the band structure of \tab \ along the symmetry lines
of the hexagonal cell. As already mentioned above, the bonding B
$\sigma$ states, which lie in the region -10 eV to -2 eV and are
highlighted in the middle panel, are completely filled. These states,
which are unoccupied along the $\Gamma$-A direction in \mgb\ (compare
to Figures 1 of Refs.~\onlinecite{pickett,kortus}), lie now between
-5~eV and -2~eV for this symmetry line and show almost 3 eV dispersion
along the hexagonal axis ($\Gamma$-A) compared to 0.6 eV in \mgb . The
two-dimensional character of these states in \mgb\ is obviously
destroyed in \tab . Furthermore, we find around the A-L-H plane of the
$k$-space a strong hybridisation of those B $\sigma$ states with the
Ta 5d$_{xz}$ and 5d$_{yz}$ states, indicated in the lower panel of
Figure 3. Once more we like to emphasize the difference to \mgb ,
where these states are of nearly pure B $\sigma$ character.
\begin{figure}[bt]
\psfig{figure=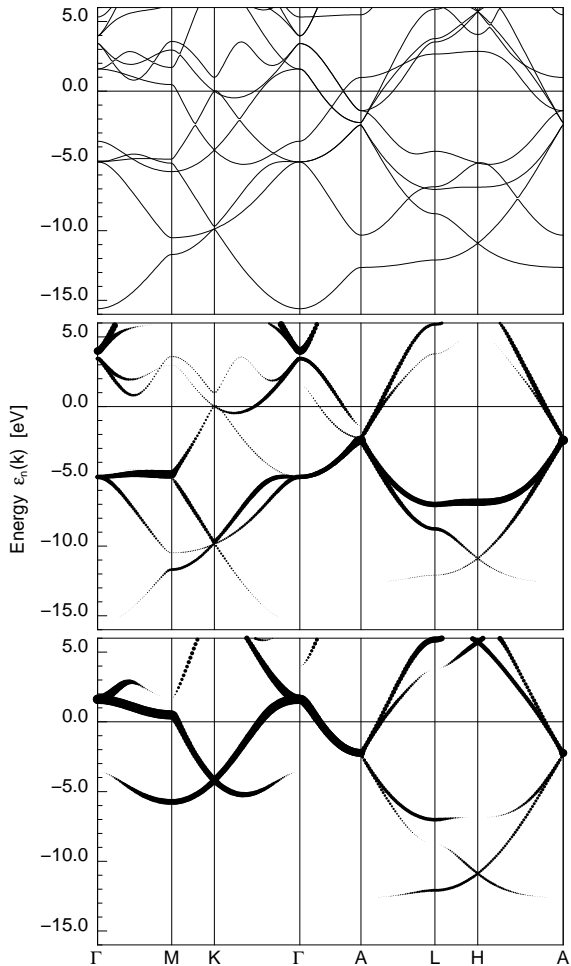,width=7.5cm}
\vspace{3mm}
\caption{ Band structure and band characters of \tab\ . The middle and
the lower panels show the band characters of the B 2p$_{x,y}$ orbitals
and the Ta 5d$_{xz,yz}$ orbitals, respectively. The line width is
scaled with the orbital weights of the corresponding orbitals.}
\label{band}
\end{figure}
The corresponding Fermi surface (FS) of \tab\ is shown in Figure
\ref{fs}. For convenience of comparison with \mgb\ (compare to Fig.~3
in Ref.~\onlinecite{kortus}), we have chosen the A point as the center
of the hexagonal prism. All three sheets of the FS of \tab\ are
electron-like. Because of the strong dispersion of the antibonding B
$\sigma$ - Ta 5d$_{xz,yz}$ states, they build closed FS around the A
point (see middle and lower panel of Fig.~4), where the hole-like
quasi two-dimensional tubes are found in \mgb . The large FS in the
upper panel of Fig.~\ref{fs} is due to Ta 5d states, the contribution
of B states to this sheet is almost negligible.
\begin{figure}[bt]
\begin{center}
\begin{minipage}{9cm}
\psfig{figure=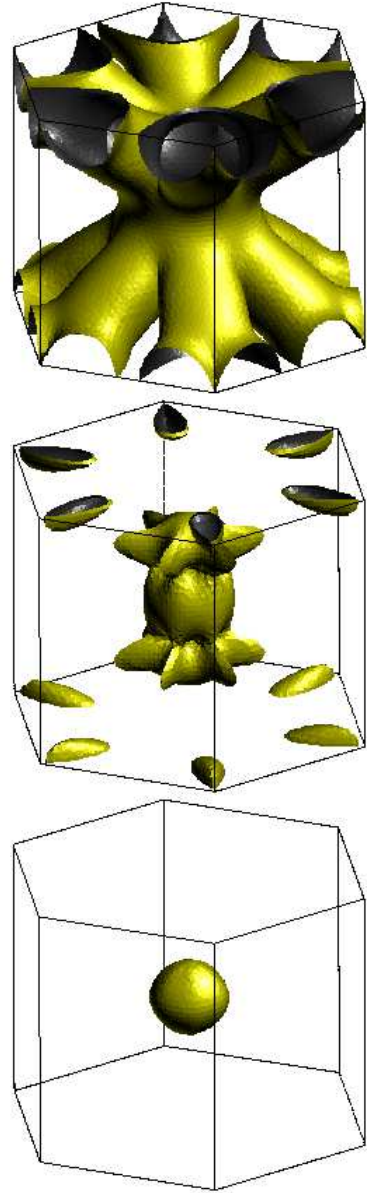,width=5.cm}
\end{minipage}
\end{center}
\vspace{3mm}
\caption{The three different sheets of the Fermi surface of \tab , all
sheets are electron-like. The
A point corresponds to the center of the hexagonal prism, the
$\Gamma$ point is the midpoint of the lower and upper hexagon.}
\label{fs}
\end{figure}

To account for the experimental uncertainty in the lattice constants
for different samples \cite{remark2}, we also investigated the
influence of different lattice constants for the experimentally
reported range on the electronic structure. The changes for the
relevant features in the band structure are negligible, the DOS is
basically unchanged, N($\varepsilon_F$) varies by less then 2\%.

For a rough estimate of the electron-phonon (el-ph) coupling in \tab ,
we calculated the phonon frequencies and the deformation potential of
the E$_{2g}$ (in-plane displacement of the borons) and the B$_{1g}$
(borons displaced along $z$ in different directions) zone-center
phonon modes. Their frequencies are 98 meV and 85 meV,
respectively. For the corresponding frequencies in \mgb\ Kortus
{\it et al.\ }\cite{kortus} reported 58 meV and 86 meV,
respectively. For AlB$_2$ 118 meV and 60 meV, respectively, were
calculated. \cite{bohnen} Already from the strong hardening of the
calculated E$_{2g}$ frequency compared with \mgb\ one can conclude a
strongly reduced electron phonon coupling of this mode.

Figure 5 shows the calculated band structure for the frozen E$_{2g}$
phonon mode of \tab\ with a B displacement of $\Delta u_B$ = 0.018 \AA
. The B bond stretching mode splits the antibonding B $\sigma$-Ta
5d$_{xz,yz}$ bands along the $\Gamma$-A line. For an averaged split
$\Delta\varepsilon_k/\Delta u_B \sim$9 eV/\AA\ we find a deformation
potential $D_{E_{2g}} \sim$ 4.5 eV/\AA\ about 3 times smaller than
$D_{E_{2g}}$ in MgB$_2$. \cite {pickett} Calculating larger elongations up to
the actual rms we found a nearly linear dependence of the deformation
potential $D_{E_{2g}}$ on the elongation $\Delta u$.

\begin{figure}[bt]
\psfig{figure=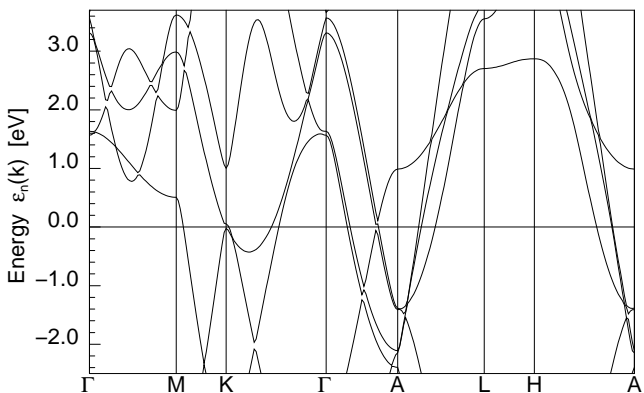,width=8.5cm,angle=-90}
\vspace{3mm}
\caption{Band structure for frozen E$_{2g}$ mode, plotted along the same lines
as in Fig.~\ref{band} to facilitate comparison. The two-fold
degenerate bands in Fig.~\ref{band}, crossing the Fermi level along
the $\Gamma$-A direction, are split due to the symmetry reducing
phonon mode.}
\label{bande2g}
\end{figure}

Following Ref.~\onlinecite{pickett}, we can estimate the coupling from
this mode alone using Eq.~(2.34) of Kahn and Allen \cite{kahn} for the
EP matrix element in terms of $D_{E_{2g}}$.
With $N_{E_{2g}}(\varepsilon_F)$ = 0.28, resulting from a summation of
the calculated orbital projected DOS for the orbitals mainly involved
in the deformed band, we find
\begin{equation}
\lambda_{E_{2g}} = N_{E_{2g}}(\varepsilon_F)
[\frac{\hbar}{2M_B\omega^2}]|\sum_{j=1,2}\hat{\varepsilon_j}D_j|^2
\sim 0.055.
\end{equation}
The sum on $j$ runs over the two B atoms of the E$_{2g}$
mode, for $M_B$ we used the isotope averaged mass of B. Due to the high
frequency $\omega$ and the smaller deformation potential $D_{E_{2g}}$,
we get a coupling smaller by a factor of about 20 compared to \mgb .

From a corresponding calculation of the coupling constant for the
B$_{1g}$ mode, reported \cite{bohnen} to be softened in AlB$_2$
($\omega$ = 60 meV), we find an even slightly smaller contribution
compared with the already weak coupled E$_{2g}$ phonon.

Assuming similar coupling to the acoustic phonons as reported
\cite{bohnen} for \mgb\ or AlB$_2$ and comparable contributions of
other modes, the total el-ph coupling constant might be no more than
$\lambda \sim$ 0.2.

\subsection{Experimental results}

\begin{figure}[bt]
\psfig{figure=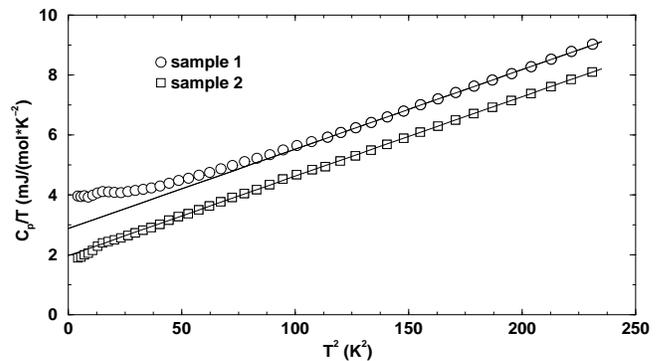,width=8.5cm,angle=-90}
\vspace{3mm}
\caption{ Specific heat $c_p(T)/T$ versus $T^2$ of two \tab\ samples
at zero magnetic field. The lines correspond to linear regression fits
of the experimental data for temperatures $T$ $>$ 8 K (sample 1) and
$T$ $>$ 4 K (sample 2), respectively.}
\label{cpfig}
\end{figure}

\begin{figure}[bt]
\psfig{figure=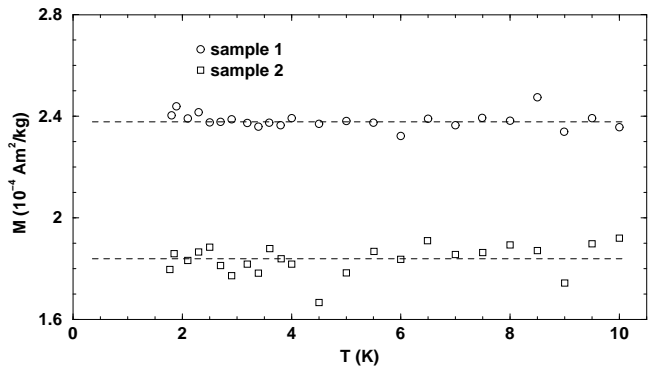,width=8.5cm,angle=-90}
\vspace{3mm}
\caption{ Temperature dependent magnetization $M$ of two \tab\ samples
in the temperature range 1.8 K $< T <$ 10 K at an applied magnetic
field of $H$ = 10 Oe.  }
\label{susfig}
\end{figure}

\begin{figure}[bt]
\psfig{figure=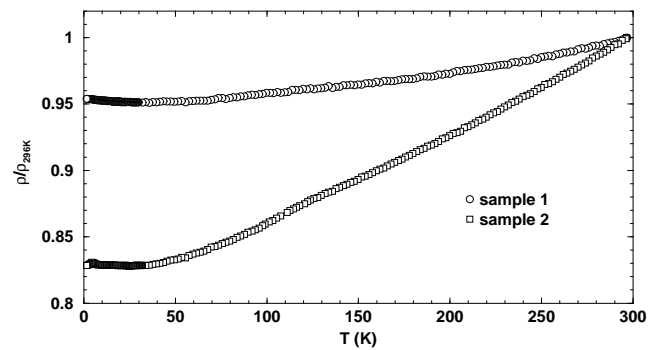,width=8.5cm,angle=-90}
\vspace{3mm}
\caption{ Temperature dependence of the resistivity normalized to its
values $\rho_{296 K}$ = 350 $\mu\Omega$cm (sample 1) and $\rho_{296
K}$ = 199 $\mu\Omega$cm (sample 2) at $T$ = 296 K in the temperature
range 1.5 K $< T <$ 296 K.  }
\label{rhofig}
\end{figure}

The results of the specific heat measurements are shown in Figure
\ref{cpfig}.  Since no superconductivity was observed at least in the
temperature range down to 1.8 K from magnetization measurements (see
Figure \ref{susfig}), and down to 1.5 K from resistance measurements
(see Figure \ref{rhofig}), the results of the specific heat
measurements are shown in the $c_p/T$ vs.\ $T^2$ plot. Thus, assuming
standard normal metal behaviour, the Sommerfeld coefficient $\gamma$
was determined using the relation
\begin{equation}
c_p/T = \gamma + \frac{12}{5}R\pi^4\Theta_0^{-3} T^2,
\end{equation}
where $R$ being the ideal gas constant and $\Theta_0$ the initial
Debye temperature.

The investigated TaB$_2$ samples show distinct anomalies at
temperatures around 4 K which are, however, not due to
superconductivity as was found from magnetization and resistance
measurements.

Sizable deviations from the linear fits, shown in Figure \ref{cpfig}
are observed especially for sample No.~1.
Similar deviations as found for these two samples are known as well
from MgB$_2$ \cite{wang,kremer,bauer}.  Since such anomalous
contributions do not essentially change in the superconducting state
of MgB$_2$, we ascribe them to lattice effects related to boron
disorder present in all diborides.

Naturally, the initial Debye temperature $\Theta_0\approx$ 417 $\pm$
2 K is smaller than the corresponding value reported for MgB$_2$ of
750 to 800 K \cite{budko,kremer,wang}. But it is somewhat {\it harder}
than the value one might expect from a simple scaling with the square
root of the total mass ratio: 369 $\pm$ 11 K. This is in line with the
hardening calculated for the optical phonons at the $\gamma$-point as
reported above.

\begin{table}
\caption{Experimental results: Specific heat (Sommerfeld constant
$\gamma$, (column 2), Debye energy $\Theta_0$ (column 3), residual
resistivity ratio RRR (column 4), and resistivity $\rho_{296K}$ at $T$
= 296 K.}
\vspace{3mm}
\begin{tabular}{c|c|c|c|c}
Sample &$\gamma$&$\Theta_0$&&$\rho_{296K}$ \\ No.\ & [mJ /mol K$^{-2}$)] & [K]&RRR&$\mu\Omega$cm \\
\hline 1 & 2.8 & 415&1.05&350\\ 2 & 2.0 & 419&1.2&199\\  \end{tabular}
\end{table}

The theoretically estimated value of the el-ph coupling constant
$\lambda$ is in accordance with those small $\lambda$ values derived by
comparing the experimental value of the Sommerfeld constant (see
figure \ref{cpfig}) with the calculated DOS at the Fermi level
$N(\varepsilon_F)$
\begin{equation}
\gamma_{exp}=(\pi^2/3)k_B^2(1+\lambda) N(\varepsilon_F)=\gamma_0
(1+\lambda).
\end{equation}
Using our calculated $\gamma_0$ = 2.14 mJ/mole$\times$K$^2$ and the
measured $\gamma_{exp}$ = 2.8 mJ/mole$\times$K$^2$ of sample No.~1, we
obtain an empirical value of $\lambda_{No.~1} = 0.3$. If this estimate
is correct, a sizable contribution of relatively low-frequency phonons
involving Ta-vibrations can be expected.

The low value of $\gamma_{exp} =2$ mJ/mole$\times$K$^2$ for sample
No.~2 can be ascribed to the significant number of boron vacancies in
understoichiometric TaB$_{2-\delta }$ samples. Within a very crude
estimate, supposing a rigid band behaviour and using the atomic
partial density of states $N_{Ta}(\varepsilon_F$) and
$N_{B}(\varepsilon_F$) (see Figure 1) we assume for the effective bare
specific heat constant $\gamma_{0,eff}$
\begin{equation}
\gamma_{0,eff} < (\pi^2/3)k_B^2(N_{Ta}(\varepsilon_F)
+(2-\delta)N_{B}(\varepsilon_F)).
\end{equation}
For TaB$_{1.29}$ (sample No.~2) this yields $\gamma_{0,eff}$ = 1.8986
corresponding to $\lambda_{No.~2}$ = 0.053.  Considering additionally
broadening effects in the DOS due to disorder in the vacancy
distribution and other impurities, that $\lambda$ value can be taken
as a lower bound.

Adopting a standard value of the Coulomb pseudopotential $\mu^*
=0.13$ one arrives at negligible values of the transition temperature
$T_c \sim 10^{-7}$ K, irrespectively of the details of the shape of
the Eliashberg function, i.e.\ the averaged phonon frequency. If even
the pseudo potential $\mu^*$ would be ignored, $T_c$ would not exceed
18 mK. Thus, the experimental evidence for the absence of
superconductivity, possibly even down to several 100
mK,\cite{samsonov} becomes very plausible already in the traditional
electron-phonon Migdal-Eliashberg picture.

\section{Conclusions}

Although \tab\ occurs in the same crystal structure as \mgb , it
should not be considered as a close relative of \mgb\ with respect to
the electronic structure (even if it would be found superconducting at
very low temperatures). The major differences occur due to: (i) the
different band filling because of the three additional valence
electrons of Ta with respect to Mg, resulting in a shift of the the
Fermi level from the bonding B-$\sigma$ states in \mgb\ to the
antibonding B-$\sigma$ - Ta-$d$ hybrid states in \tab , (ii) the
strong out of plane hybridization of the B 2$p$ states with Ta, (iii)
the weak electron phonon coupling, especially of the E$_{2g}$ mode,
which is strongly coupled in \mgb .

According to our experimental results, TaB$_2$ is not superconducting
down to 1.5 K. Thus, we confirm the earlier results of Refs.\
\cite{gasparov,samsonov} and disprove at the same time the speculation
about superconductivity around 9.5 K reported in Ref.\
\cite{kaczorowski1}.  Because in our opinion the main reason for the
absence of superconductivity in \tab\ is the different position of the
Fermi level (with respect to MgB$_2$), huge hole doping might
``reintroduce'' superconductivity at relatively high temperatures.

In other words, the results obtained here suggest that the empirical
absence or low-temperature superconductivity established in many
transition metal (or rare earth) diborides with {\it electrons} as the
potentially paired charge carriers, stressed by Hirsch \cite{hirsch},
might be explained in the traditional electron-phonon picture simply
by a {\it weak} electron-phonon interaction. In that case there is no
need to explain this behavior by the absence of a sophisticated
Coulomb interaction driven non-phonon mechanism which works
exclusively for holes\cite{hirsch}.

Corresponding studies for other related transition metal diborides of
experimental interest will be published elsewhere.

\section{acknowledgments}

We thank J.M.\ An and S.V.\ Shulga for discussions and N.\ Mattern for
providing us with the XRD-data.  This work was supported by the DAAD
(individual grant H.R.), the ONR Grant No.\ N00017-97-1-0956, the SFB
463, and the Deutsche Forschungsgemeinschaft.


\begin{references}
\bibitem{remark1} Our band structure calculations for MgB$_2$ are in
excellent agreement with those of
Refs.~\onlinecite{pickett,kortus}. The partial DOS naturally differ
from Ref.~\onlinecite{kortus} because of different calculational
methods. To calculate the partial DOS, we use the projection onto our
minimal basis set instead of a projection onto muffin tin spheres in
Ref.~\onlinecite{kortus}, which finds a large contribution of the
interstitial plane wave region.
\bibitem{remark2} The published lattice constants for \tab\ scatter
much stronger than the usual experimental errors in an XRD
experiment. We found parameter sets reported with 3.057 \AA\ $<$ a $<$
3.0966 \AA\ and 3.224 \AA\ $<$ c $<$
3.288.\cite{pearson,gasparov,japaner} The different lattice constants
are most probably due to different B contents of different samples.
\end{references}
\end{document}